\documentclass[prl,aps,groupedaddress,superscriptaddress,notitlepage, twocolumn]{revtex4-2}

\usepackage[utf8x]{inputenc}
\usepackage{color}
\usepackage{bbm} 

\usepackage{nicefrac}
\usepackage{amsfonts,amsmath,amssymb,stmaryrd}
\usepackage{braket}

\usepackage[autostyle=true]{csquotes}

\usepackage{tabularx}
\usepackage{multirow} 
\usepackage{hhline}
\usepackage{graphicx}
\usepackage{subfigure}  
\usepackage{bbm} 
\usepackage{hyperref}
\usepackage[pdftex]{epsfig}
\usepackage{mathrsfs}
\usepackage{verbatim}
\usepackage{centernot}
\usepackage{array}
\usepackage{cancel}
\usepackage{ifthen}
\usepackage{bm}	
\usepackage{siunitx}
\usepackage{float} 
\usepackage[super]{nth} 
\usepackage{todonotes}
\usepackage{color,soul}
\InputIfFileExists{gitinfo-latex.inc}{}{}

\usepackage{listings}
\usepackage{color}
\usepackage{textcomp}

\usepackage[acronym,nomain]{glossaries}

\begin{document}
	
\title{Coherent and dephasing spectroscopy for single-impurity probing of an ultracold bath}
\author{Daniel Adam}
\affiliation{Department of Physics and Research Center OPTIMAS, Technische Universit\"at Kaiserslautern, Germany}

\author{Quentin Bouton}
\affiliation{Department of Physics and Research Center OPTIMAS, Technische Universit\"at Kaiserslautern, Germany}

\author{Jens Nettersheim}
\affiliation{Department of Physics and Research Center OPTIMAS, Technische Universit\"at Kaiserslautern, Germany}

\author{Sabrina Burgardt}
\affiliation{Department of Physics and Research Center OPTIMAS, Technische Universit\"at Kaiserslautern, Germany}

\author{Artur Widera}
\email{email: widera@physik.uni-kl.de}
\affiliation{Department of Physics and Research Center OPTIMAS, Technische Universit\"at Kaiserslautern, Germany}
\date{\today}

\begin{abstract}
We report Ramsey spectroscopy on the clock states of individual Cs impurities immersed in an ultracold Rb bath. 
We record both the interaction-driven phase evolution and the  decay of fringe contrast of the Ramsey interference signal to obtain information about bath density or temperature nondestructively.  
The Ramsey fringe is modified by a differential shift of the collisional energy when the two Cs states superposed interact with the Rb bath.  
This differential shift is directly affected by the mean gas density and the details of the Rb-Cs interspecies scattering length, affecting the phase evolution and the contrast of the Ramsey signal.  
Additionally, we enhance the temperature dependence of the phase shift preparing the system close to a low-magnetic-field Feshbach resonance where the $s$-wave scattering length is significantly affected by the collisional (kinetic) energy. 
Analyzing coherent phase evolution and decay of the Ramsey fringe contrast, we probe the Rb cloud's density and temperature. 
Our results point at using individual impurity atoms as nondestructive quantum probes in complex quantum systems.

\end{abstract}

\maketitle

\subsection{}
Individual impurities immersed in a gas form a paradigm of open quantum systems. 
An application of this scenario, which has attracted significant interest in recent years, is quantum probing, where information of a many-body system is mapped nondestructively onto quantum states of the impurity. 
A prominent example is thermometry of quantum gases, where in the ultracold domain, precise temperature information in the nanokelvin range is to be determined.  
Realizations of impurity-based quantum gas thermometry include mapping  thermal information onto the classical motional state of ensembles of impurities \cite{Spiegelhalder2009,Olf15,Lous2017}, single impurities \cite{Hohmann2016} or the impuritie's quantum spin distribution \cite{Bouton2020}. These methods, however, either used classical degrees of freedom or inelastic processes, perturbing the many-body system either by exchange of energy or angular momentum.
In order to tackle this situation, advanced proposals suggested using coherent superposition of quasi-spin states of single atoms to store information about the gas \cite{Ng08,Klein_2007}.  In the context of many-body physics the spin coherence of neutral impurities in an ultracold gas has been studied 
\cite{Edri2020, Cetina2016}. 
Besides, recent works predict enhancing the performance of thermometry by exploiting specific properties of the bath, which modify the nonequilibrium dynamics of the probe \cite{Mehboudi2019}.  
For a Bose-Einstein condensate, an impurity can form a polaronic quasi-particle,  and thermal information can be obtained from  fluctuations of the probes' momentum and position \cite{Mehboudi19}. By contrast, in a Fermi gas the existence of a Fermi sea allows deducing thermal information from the dephasing dynamics of a coherent superposition of internal probe states \cite{Mitchison2020}.

\begin{figure}
\includegraphics[width=0.45\textwidth]{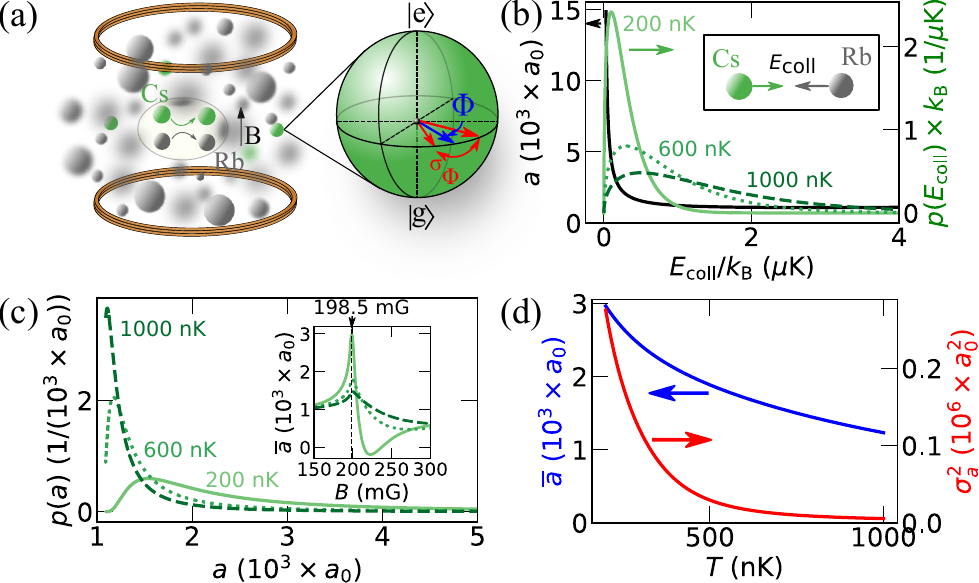}
\caption{(a) Single impurity (green) immersed in Rb bath (gray) interacting. The coherent superposition of Cs states is visualized on a Bloch-sphere showing the interaction-induced phase evolution $\Phi$ and phase dispersion $\sigma_\Phi$. 
	(b) Scattering length $a$ (black solid) for the Cs ground state $\ket{g}$ interacting with the Rb bath for different collision energies (inset) in the vicinity of a Feshbach-resonance at 198.5 mG. On the same graph, we also plot the Maxwell-Boltzmann distribution of collision energies in green for 200 nK, 600 nK, and 1000 nK from light to dark green in solid, dotted, and dashed lines, respectively.
	(c) Probability distribution of the scattering length $a$ for different bath temperatures; colors and line styles as in (b). The inset shows a temperature averaged scattering length $\overline{a}$ versus $B$-field. The black dashed line marks the experimentally fixed magnetic field $B$ at $ 198.50 \pm 0.05 $ mG, at the peak of the Feshbach resonance. 
	(d) Mean scattering length $\overline{a}$ (blue) as well as the variance of the probability distribution of scattering length $\sigma_a^2$ (red) as a function of temperature. 
}
\label{fig:setup}
\end{figure}

Here, we couple individual Cs atoms in a coherent superposition of the clock states to an ultracold bath of Rb atoms, see Fig. \ref{fig:setup}.  
We prepare the gas just above the critical temperature for the condensate, with temperatures in the range $T/T_c = 1.2 \ldots 5$, with $T_c$ the critical temperature for condensation.  
Despite a relatively large interspecies $s$-wave scattering length of several thousand Bohr radii, see Fig.~\ref{fig:setup} (b), our configuration leads to a Cs mean free path that is similar or larger than the Cs impurity's de Broglie wavelength.
Therefore our system differs from the recently reported Bose-polarons \cite{Yan190,Guenther18, Liu19,Dzsotjan20, Skou21}. 
However, comprising an impurity in a bosonic bath just above the condensation threshold, it bears similarities with impurities immersed in an ultracold Fermi gas \cite{Cetina2016,Schmidt_2018}, as many wave vectors contribute to collisions.
Using Ramsey spectroscopy on the Cs clock states we monitor both the coherent interaction-induced frequency shift on the coherent superposition and the nonequilibrium decoherence of the probes.  
Thanks to comparison to a microscopic model, including precise knowledge of the molecular two-body interaction potential, we demonstrate the capability for single-atom quantum probing via coherent or dephasing signals in ultracold gases. 

The microscopic mechanism of our probing scheme is illustrated in Fig.~\ref{fig:setup}(b-d).
In the regime of ultracold temperatures probed in this work, the interaction strength between Rb and Cs is only described by the $s$-wave scattering length $a$.  For a given scattering length $a_i$ between the impurity state $\vert i\rangle$ and the Rb cloud, the interaction energy writes $ E_{i} = 2\pi \hbar^2 n a_i/\mu$, where $n$ is the Rb density, $\mu$ the reduced mass, and $\hbar$ the reduced Planck constant. 
It leads to a time-dependent phase shift of the coherent superposition $(\ket{g}+ie^{i\Phi(t)} \ket{e})/\sqrt{2}$ of the probe with phase \cite{Schmidt2018}
\begin{align}\label{eq:detuning}
\Phi (t) = \delta_\mathrm{Rb}  t = \frac{2\pi \hbar^2 n \Delta a}{\mu \hbar} t,
\end{align}
where $\Delta a$ is the difference of scattering lengths between excited and ground states $\Delta a = a_{e}-a_{g}$, leading to an interaction-induced energy shift $\hbar \delta_\mathrm{Rb}$ between the two impurity clock states.  
The information connected to density and temperature in this phase shift can be inferred from the phase and decoherence of the Ramsey signal after interaction time $t$ as explained below.
We enhance the sensitivity of the Ramsey phase to temperature by tuning the scattering length $a_g$ close to the maximum of an interspecies Feshbach resonance at a magnetic field $B_0=198.50\pm 0.05\,$mG. By contrast, the scattering length $a_e$ is only slightly changing throughout this work and we assume $a_e$ to be constant in the following.
At the low magnetic field value close to the resonance, $a_g$ changes significantly for changing collisional, i.e. kinetic, energy in an interatomic collision, see Fig.~\ref{fig:setup}(b).  
For a thermal ensemble with a fixed magnetic field close to the resonance, thermal averaging leads to a temperature-dependent distribution of scattering lengths, see Fig.~\ref{fig:setup}(c) and Eq.~\ref{eq:MBD} in \cite{Supplementary}. 
The resulting mean scattering length, see Fig.~\ref{fig:setup}(d), drives the evolution of the interaction-induced phase shift of the impurity's superposition, which can be read out from our Ramsey-type scheme.   
Broader distributions cause stronger dispersion of the Bloch vector and hence dephasing,  reflected by a decreasing contrast in a Ramsey-fringe measurement. This can be seen in analogy with interference from a spectrally broad light source in optics.  Thereby, the Ramsey-fringe signals allow determining thermal properties of the bath from coherent evolution and dephasing dynamics of a single-atom probe.  
Additionally, the inhomogeneous density profile, described by a Maxwell-Boltzmann distribution, leads to a spatial dependence of the phase shift of the probe.
As a consequence, the phase evolution of the impurity evolves faster in the center of the Rb cloud than in the wings.
Since the width of the density distribution changes with temperature, the dephasing signal and phase shift yield information on the atomic density.
For fixed density, the comparison between mean value and the width of the scattering-length distribution allows for distinguishing two cases. 
First, for widths much smaller than the mean scattering length, dephasing emerges on a time scale longer than the inverse angular frequency of the Bloch vector precession on the Bloch sphere equator.  
In this configuration, the Ramsey-type interference signal will show a coherent phase shift of the Bloch vector. 
By contrast, for large widths of the scattering length distribution, the Bloch vector  dephases before completing a single revolution. 
In this case, the coherent phase shift cannot be reliably detected, but equivalent information can be extracted from the dephasing. 
Fig.~\ref{fig:setup} (d) illustrates these two situations. At large temperature, the Ramsey signal is dominated by the by the coherent phase shift, while at low temperature, dephasing is prevailing.
We note that this unusual situation arises due to the proximity to the Feshbach resonance.
\\

\begin{figure}
	\includegraphics[width=0.45\textwidth]{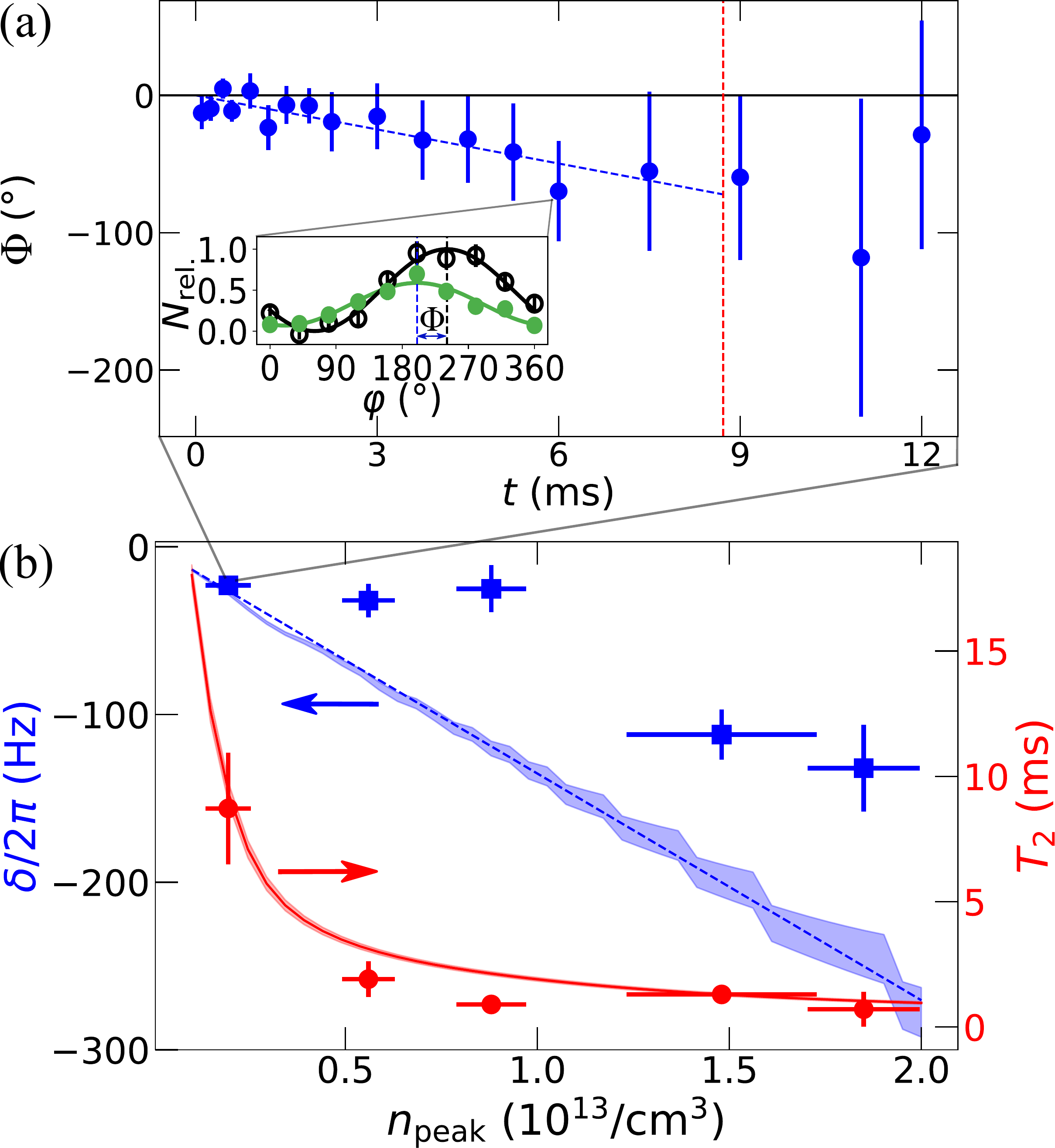}
	\caption{(a) Interaction-induced phase shift $\Phi(t)$ as a function of free-evolution time $t$. The data is fitted with a line (dashed blue line) to extract the slope $\delta$ up to the dephasing time $T_2$, indicated by the vertical dashed line.  The inset shows a Ramsey fringe for $t=6\,$ms with (green, filled circles) and without (black, open circles) Rb cloud.  (b) Measured interaction-induced detuning $\delta$  (blue data points) and $T_2$ times (red data points) as a function of gas density. The red solid and blue dashed lines represent our simulation results, see \cite{Supplementary}. The fixed temperature $T=850\,$nK with changing density leads, in this dataset, to a range of $T/T_c=2.2\ldots5$ for the Rb gas.
    } 
	\label{fig:density}
\end{figure}

Experimentally, we prepare up to ten Cs impurities in the hyperfine ground state $\ket{F_\mathrm{Cs} = 3, m_{F,\mathrm{Cs}} = 0}$ at a temperature of $T_\mathrm{Cs}=1.7\,\mu$K and immerse them in a Rb bath. The Rb bath is prepared in state $\ket{F_\mathrm{Rb}=1, m_{F,\mathrm{Rb}} = 1}$, where $F_j$ and $m_{F, j}$ are the total atomic angular momentum and its projection onto the quantization axis for species $j=$ Cs or Rb, respectively. The Rb cloud is produced with temperatures and peak densities in the range of $T = 200\ldots 1000$ nK and $n_0 = 0.2\times10^{13} \ldots 2\times10^{13}$ cm$^{-3}$, respectively \cite{Supplementary}.
\begin{figure}
	\includegraphics[width=0.45\textwidth]{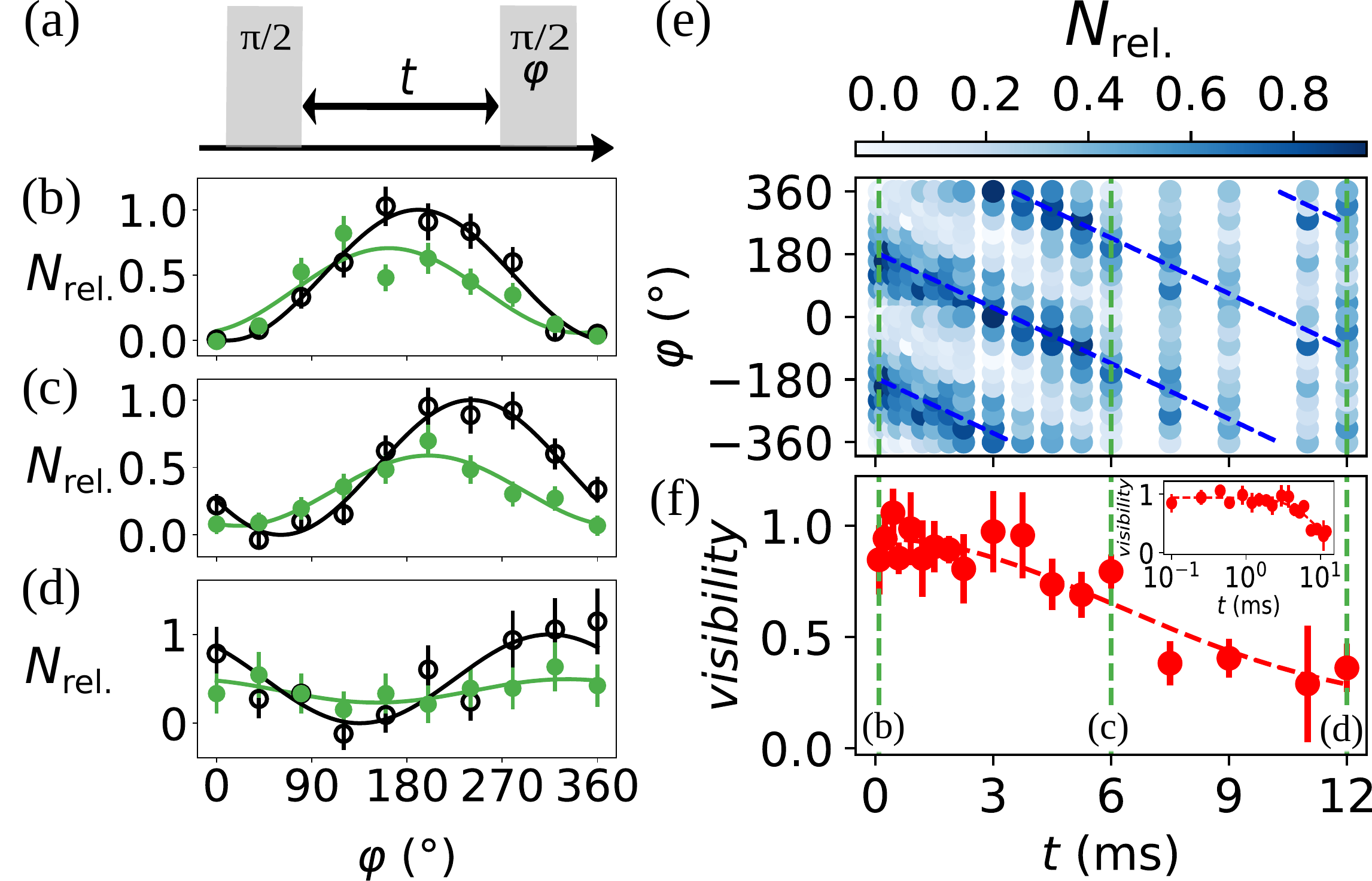}
	\caption{ (a) In our Ramsey sequence,  a first $\pi/2$-pulse is followed by a free evolution time $t$. The final pulse has an adjustable phase  $\varphi$  between 0\textdegree{} and 360\textdegree{} to reveal the Ramsey fringe. The Ramsey fringes are shown for (b) $t=0.1\,$ms, (c) $t=6\,$ms, and (d) $t=12\,$ms for the case without (black, open circles) and with (green, filled circles) Rb bath. 
		(e) Time evolution of the color-coded Ramsey signal for a bath peak density of $n_\mathrm{Rb}= 0.195\times10^3\,$cm$^{-3}$ and temperature of $T=980\,$nK, showing the phase evolution.  The phase shift is directly proportional to evolution time, where the slope is given by the energy difference between the two Cs clock states.  The dashed line is a linear fit to the fringe maxima; the negative phase is a copy of positive phase data for better illustration.
		(f) Decay of Ramsey fringe contrast with time.  The decoherence time $T_2$ is fitted by a Gaussian function. The inset shows the same data on a logarithmic time scale.
	}
	\label{fig:data}
\end{figure}
Thereafter, the Ramsey sequence (see Fig.~\ref{fig:data}(a)) is initialized on the clock transition $\ket{g} = \ket{F_\mathrm{Cs}=3, m_{F,\mathrm{Cs}} = 0} \rightarrow \ket{e} = \ket{F_\mathrm{Cs}=4, m_{F,\mathrm{Cs}} = 0}$ by a first microwave $\pi/2$-pulse with Rabi frequency $\Omega_0 = 2\pi \times 15.4$ kHz, preparing the coherent state $\ket{\psi(0)} = \frac{1}{\sqrt{2}}(\ket{g}+i \ket{e})$. 
During a free evolution time $t$ the superposition acquires a phase $\ket{\psi(t)} = \frac{1}{\sqrt{2}}(\ket{g}+i e^{-i \Delta\times t} \ket{e})$, where $\hbar \Delta$ is the total energy difference between the Cs clock states. It contains differential light shifts and the second order Zeeman shift, as well as the interaction-induced shift $\hbar \delta_\mathrm{Rb}$.
After an interaction time $t$ between the impurity and the bath, a second $\pi/2$-pulse is applied with adjustable phase $\varphi$ relative to the first pulse.
To obtain the typical Ramsey fringe at a fixed time $t$ (see Fig.~\ref{fig:data} (b-d)), we measure the population in state $\ket{g}$, see \cite{Supplementary}, for the full phase range of $\varphi = 0$\textdegree $\ldots 360$\textdegree. Fig.~\ref{fig:data} (e) shows the full time evolution.
A species-selective lattice allows obtaining spatial resolution along the axial direction to select those Cs atoms spatially overlapping with the Rb cloud. 
The measured population in $\ket{F=3}$ after the Ramsey sequence is modeled by 
\begin{align}\label{eq:FringeDephasing}
 p(t, \varphi)= \frac{1}{2}+\left( \sin^2\left[ \frac{\Delta \cdot t-\varphi}{2} \right] -\frac{1}{2} \right)\exp\left[-\frac{t^2}{T_2^2}\right],
\end{align}
where we have added phenomenological Gaussian dephasing with dephasing time $T_2$.
In order to analyze the data, we normalize each fringe to obtain relative numbers (see \cite{Supplementary}) and fit the fringes with a function $A \sin^2 [ (\phi-\varphi)/2 ]+C$, with amplitude $A$, offset $C$, and fringe phase $\phi$. From amplitude and offset we deduce the fringe visibility, for each $t$, as ${\cal{V}}=(p_\mathrm{max}-p_\mathrm{min})/(p_\mathrm{max}+p_\mathrm{min})=A/(A+2C)$. From difference in contrast for measurements with and without Rb cloud, we determine the interaction-induced visibility loss as a measure of dephasing, see Fig.~\ref{fig:data} (f). The $T_2$ time is now extracted with a Gaussian fit to the visibility
\begin{align}
{\cal{V}}(t)={\cal{V}}_0\exp \left (- \frac{t^2}{T_2^2} \right )+B,
\label{eq:visibility}
\end{align}
with amplitude ${\cal{V}}_0$ and offset $B$.  The offset is due to residual Cs atoms non interacting with the Rb cloud
To extract only the interaction-induced phase for every free-evolution time $t$ (see Fig.~\ref{fig:density}(a)), we subtract the contributions of interaction with the environment besides the bath interaction, as $\Phi(t) = \phi(t)-\delta_\mathrm{bg} t$, with $\delta_\mathrm{bg}$ the phase shift without the Rb bath.
The dominating contributions to $\delta_\mathrm{bg}$ are the differential light shift $\delta_\mathrm{DT}$ of the dipole trap and the second-order Zeeman shift $\delta_B$ \cite{Supplementary}. 
We find with an independent characterization $\delta_\mathrm{bg} = \delta_\mathrm{DT} + \delta_B \approx -2\pi\times 135 $ Hz for a Cs temperature of  $T_\mathrm{Cs} = 1.7\,\mu$K. We use this value throughout the analysis, neglecting small changes of the detuning for reduced temperatures of the Cs atoms \cite{Kuhr2005}. 
For further characterization we also extract the dephasing $T_2$ time without Rb, originating from the harmonic dipole trap potential \cite{Kuhr2005}. We find $T_2 = 27.2$\,ms, which is much longer than the dephasing time when the impurity is immersed in the Rb bath. Therefore, this timescale will be neglected in the following.  
Besides, decoherence originating from decay to the ground state, quantified as $T_1$ time, is, in our case, absent on all relevant time scales, $ T_1 \gg T_2 $, and therefore also neglected.\\

We first probe the gas density information at a relatively high temperature of $T=850\,$nK, as shown in Fig.~\ref{fig:density}, where the phase dispersion is sufficiently small to monitor the phase evolution \cite{Supplementary}. We record Ramsey fringes for variable interaction times with and without a Rb bath and extract the interaction-induced phase shift dynamics and contrast decay.  For a given parameter set, the phase difference shows a linear change with time, where the slope is given by the interaction-induced detuning $\delta_\mathrm{Rb}$. 
In addition we extract the dephasing time $T_2$ as in Eq.~\ref{eq:visibility}.
For the dephasing times $T_2$ extracted, we find a reduced dephasing time $T_2$ for increasing density as shown in Fig.~\ref{fig:density}(b).
We compare our data to a model without free parameters, computing Ramsey fringes taking into account the Rb density distribution as well as the temperature-dependent distribution of  $s$-wave scattering lengths, see Eq.~(\ref{eq:DephasingModel}) in Ref.~\cite{Supplementary}.  We analyze the numerical signals in the same way as the experimental ones to obtain the phase shift $\delta$ and dephasing time $T_2$. 
While the phase evolution shows qualitatively similar behavior between our experimental data and our model, but quantitative differences, we find good agreement for the dephasing. This supports our assumption that the Rb bath's inhomogeneous density dominates the dephasing at this temperature. We emphasize that both theory lines in Fig.~\ref{fig:density}(b) originate from the same parameter-free model Eq.~(\ref{eq:DephasingModel}).  We conclude that phase shift and decoherence both reflect the atomic peak density.

\begin{figure}
	\includegraphics[width=0.5\textwidth]{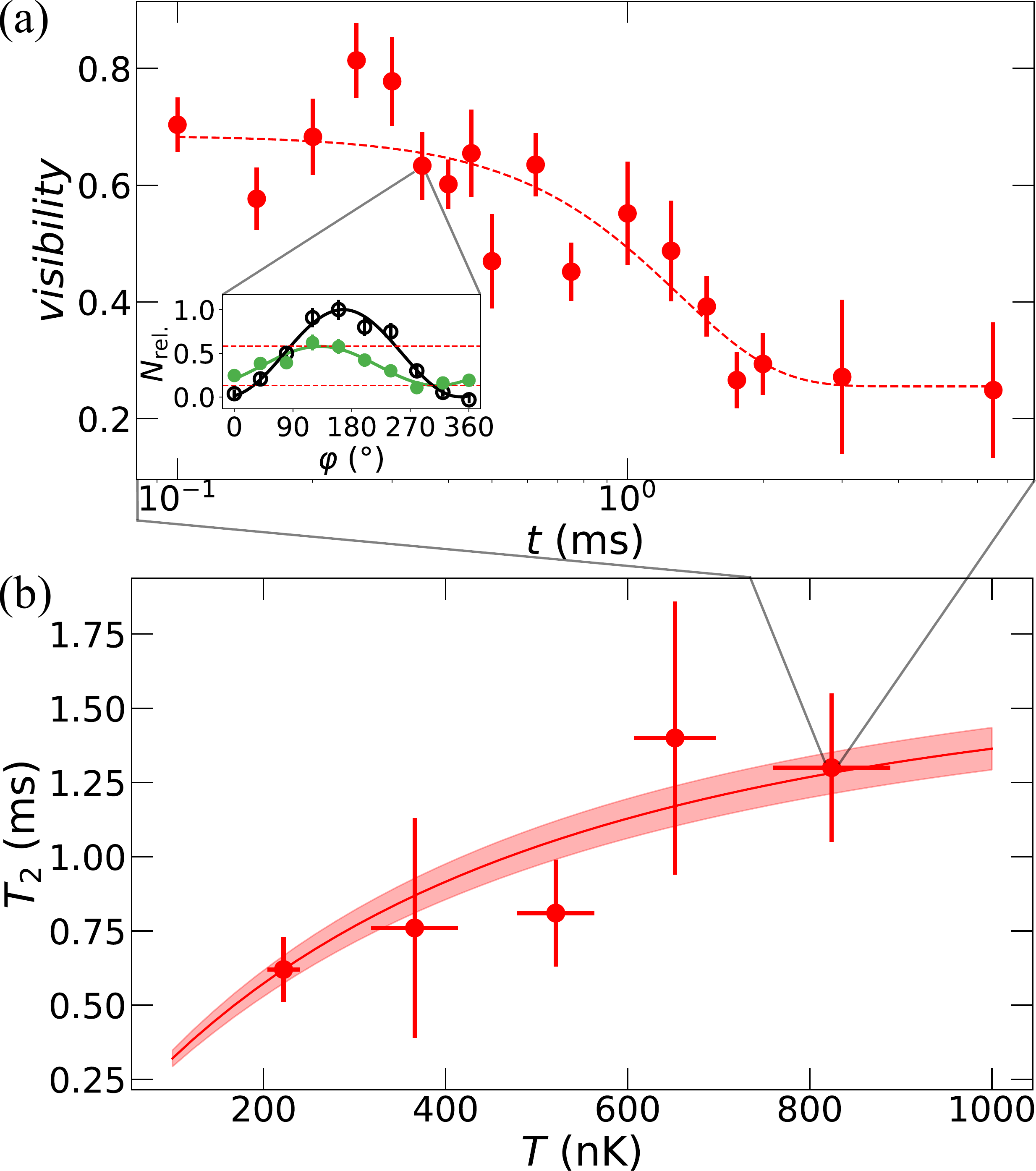}
	\caption{(a) Decay of the Ramsey-fringe contrast  as a function of evolution time in the Rb bath. The dashed line is a Gaussian fit. Inset shows Ramsey fringes for $t=0.6\,$ms for Cs atom without (black, open) and with (green, filled) Rb bath present.  (b) Dephasing time of impurities in the Rb bath as a function of gas temperature. The data (red points) show an increase with temperature; the red line is a prediction of our model. Data are taken for a Rb peak density of $n_\mathrm{Rb} \approx 1.5\times 10^{13}$ cm$^{-3}$.  Error bars and shaded area indicate $1\,\sigma$ uncertainties of the experimental data and numerical model, respectively. Scaled with the critical temperature, the temperature range is $T/T_c = 1.2 \ldots 2.3$.}
	\label{fig:temperature}
\end{figure}
Second, we study the nonequilibrium dephasing of the coherent superposition as a probe for the temperature of the Rb cloud, following the spirit of Ref.~\cite{Mitchison2020}. 
When probing the temperature of the gas, we find that the phase shift cannot be reliably extracted from the data for low temperatures, as expected. 
Instead, we use the decoherence and measure the characteristic coherence time $T_2$ to obtain information about the temperature of the gas and compare it with the independently measured gas temperature via time-of-flight velocimetry. 
To this end, we measure Ramsey signals of individual Cs impurities immersed into an atomic bath of density  $n(0) \approx 1.5\times 10^{13}$ cm$^{-3}$. 
Fig.~\ref{fig:temperature} shows the extracted $T_2$ time for a given temperature, with the visibility fitted as in Eq.~(\ref{eq:visibility}).

We find a good agreement between the experimental data and our parameter-free model. We emphasize here that the temperature axis in Fig.~\ref{fig:temperature}(b) indicates time-of-flight temperatures of the gas, where the numerical model has assumed thermalization of the Cs impurity with the cloud.  Thus, our results demonstrate that temperature information can be obtained from the dephasing Ramsey signal of individual impurities coupled to a gas.
 \\

The ability to use the coherent and decoherence dynamics of the superposition states of a single-impurity probe brings probing of a many-body system to the quantum level. 
An interesting question concerns the degree of perturbation of the measurement on the many-body system. For a Bose-Einstein condensate in a single quantum state, the probe's phase evolution corresponds to an entanglement of the condensate properties with the impurity atom.  A partial measurement of the impurity state would realize a measurement at the Heisenberg limit, perturbing the quantum gas to the smallest degree possible. For the thermal gas considered here, the effect of the impurity collisions onto the Rb system is harder to visualize.  Nominally, the number of collisions for Cs state $\ket{g}$ is in the range of $6\ldots18$ during the $T_2$ time, whereas the number of collisions in state $\ket{e}$ is in the range of $0.4\ldots1.8$ during the $T_2$ time, reflecting the strongly differing scattering cross sections.  However, the overall perturbation of the Rb gas is at the level of perturbation dictated by quantum physics.
Our work thereby paves the way to experimental probing of a qubit coupled to an open quantum system. In the future,  it will be interesting to study the decoherence dynamics originating from collisional dephasing in engineered baths with, e.g., reduced dimensions, bath-spin degree of freedom, or nonequilibrium states.

\textbf{Acknowledgements--- }
We thank E.~Tiemann for providing us with the energy-dependent scattering lengths for  Rb-Cs collisions. We acknowledge helpful discussions with Thomas Busch and Miguel Garcia-March, which helped us improving the presentation.
This work was supported by Deutsche Forschungsgemeinschaft (DFG) via Sonderforschungsbereich SFB/TRR 185, project number 
 277625399.

\bibliography{bibliography}

\newpage

\newpage
\section*{SUPPLEMENTARY}

\subsection{Overview of the experiment}

\begin{figure}
	\includegraphics[width=0.35\textwidth]{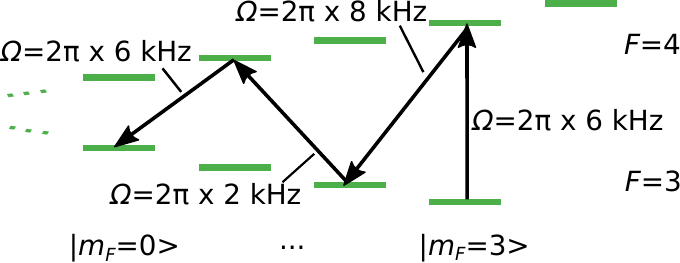}
	\caption{Preparation of the Cs state $\ket{F=3, m_{F}=0}$. After initial preparation in $\ket{F=3, m_{F}=3}$ a series of Landau-Zener sweeps is transferring the Cs atoms into the ground state of the Ramsey spectroscopy. The transfer efficiency from state $\ket{m_{F}=3}$ to $\ket{m_{F}=0}$ is $\approx 45$\%. After the preparation steps we have up to 15 atoms, of which ten atoms are in state $\ket{m_{F}=0}$.}
	\label{fig:LZ sweeps}
\end{figure}

We prepare up to ten Cs impurity atoms in the hyperfine ground state $\ket{F_\mathrm{Cs} = 3}$.
The atoms are cooled down to 1.7 µK by Raman-sideband cooling \cite{Mayer2020}, transferring the atoms into the lowest energy Zeeman-state $\ket{F_\mathrm{Cs} = 3, m_{F,\mathrm{Cs}} = 3}$. 
The temperature of the Cs atoms is measured by recording the cumulated energy distribution through the adiabatic lowering of the trap depth \cite{Alt03, Supplementary}. 
The transfer to the initial state of the Ramsey spectroscopy $\ket{F_\mathrm{Cs} = 3, m_{F,\mathrm{Cs}} = 0}$ is performed with a series of Landau-Zener sweeps starting with the Cs atoms in $\ket{F_\mathrm{Cs}=3, m_{F,\mathrm{Cs}} = 3}$ of the Raman cooling.
Fig.\ref{fig:LZ sweeps} shows the order of the sweeps indicated by arrows. This order provides a fast transfer, optimized for a minimum number of pulses and avoiding sweeps with the lowest Rabi frequency, here the $\sigma^-$ transitions. This preparation is followed by a resonant laser pulse resonant to the $\ket{F_\mathrm{Cs}=4}\rightarrow \ket{F_\mathrm{Cs}’=5}$ transition and a cleaning scheme, based on the spin selective readout of \cite{Schmidt2018}, to remove residual atoms not in $\ket{F_\mathrm{Cs}=3, m_{F,\mathrm{Cs}} = 0}$. Due to technical imperfections,  a few atoms may populate other than the target state leading to an offset in the Ramsey fringes. However, this offset will not change the interpretation of the Ramsey fringes.
The Rb bath is independently prepared in the hyperfine state $\ket{F_\mathrm{Rb}=1, m_{F,\mathrm{Rb}} = 1}$ spatially separated from Cs by ~200 $\mu$m. 
The configuration of internal states suppresses inelastic spin-exchange collisions \cite{Schmidt2019}, and only elastic collisions between bath and impurity are allowed. 
The Rb gas can be prepared in a wide range of temperatures $T = 200\ldots 1000$ nK and peak densities $n_0 = 0.2\times10^{13} \ldots 2\times10^{13}$ cm$^{-3}$. 
After the preparation step, the impurity is transported into the bath using a species-selective optical lattice \cite{Schmidt2016}, and we allow for thermalization during a waiting time of $1.4\,$ms corresponding to one complete oscillation in the trap with trap frequency $\omega_{r}\approx 2\pi \times 681\,$Hz.
For the read-out of Cs atoms after the Ramsey scheme another resonant pulse resonant to the $\ket{F_\mathrm{Cs}=4}\rightarrow \ket{F_\mathrm{Cs}’=5}$ transition is applied. We take a fluorescence image of the remaining atoms, i.e., counting only atoms in $\ket{F_\mathrm{Cs}=3}$.

\begin{figure}\label{fig:BareFringes}
	\includegraphics[width=0.5\textwidth]{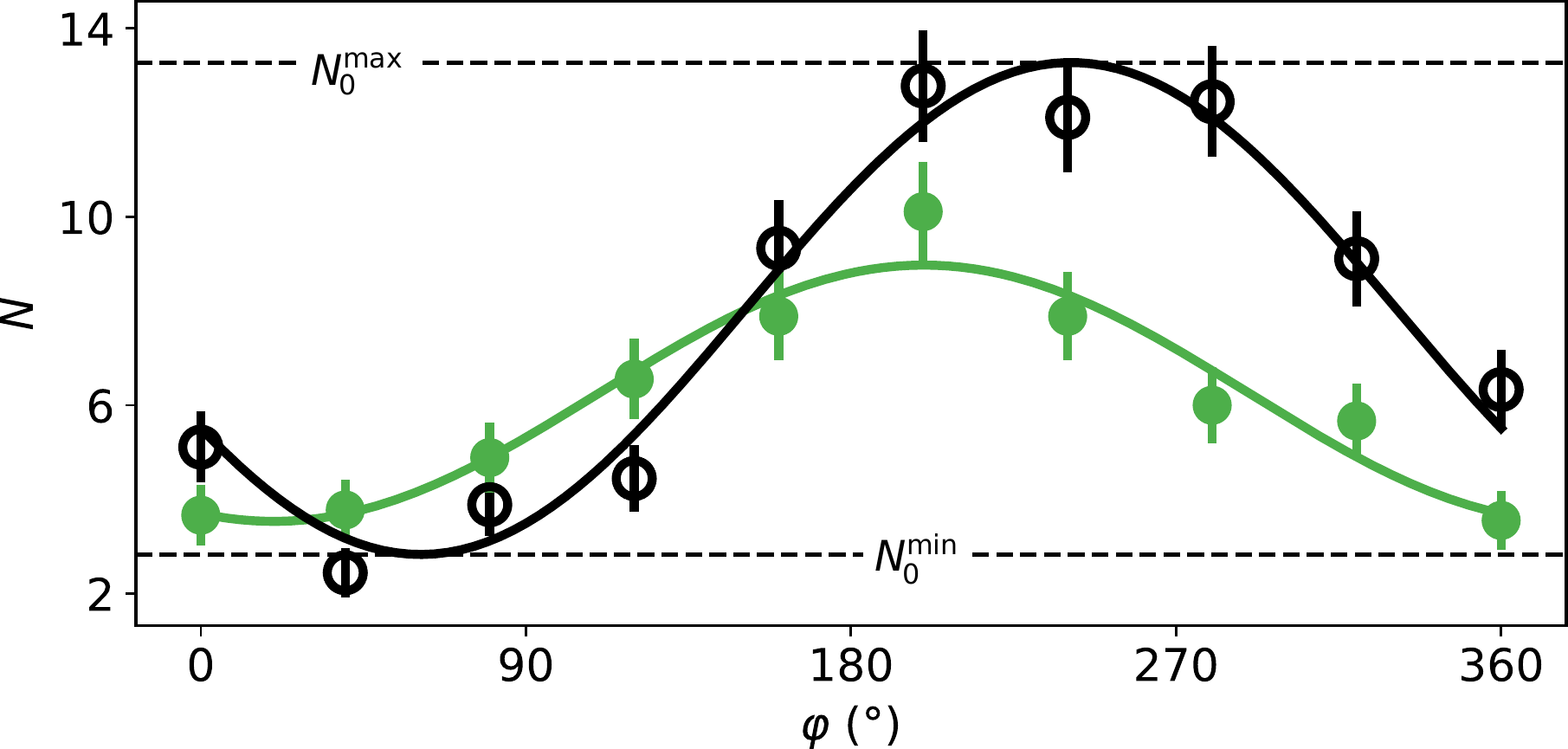}
	\caption{The Ramsey fringe with total Cs atom number as shown in FIG.~ \ref{fig:data} (c).}
	\label{fig:total Cs atom number}
\end{figure}

We calculate Ramsey signal of the relative atom number $N_\mathrm{rel}$ from the measured fringes with (measuring $N_\mathrm{bath}$ atoms) and without Rb bath (measuring $N_0$ atoms)  present, see Fig.~\ref{fig:BareFringes}. The maximum atom number $N_0^\mathrm{max}$ of the bath-free fringe sets the scaling factor for normalization. The minimum atom number $N_0^\mathrm{min}$ of the bath-free fringe indicates a fraction of Cs atoms that are not in the $m_F=0$ state and do not contribute to the Ramsey signal. Therefore, we first subtract the fringe offset without Rb from the fringe data with Rb bath present and then scale the data for the maximum atom number without Rb
\begin{align}
N_\mathrm{rel} = \frac{N_\mathrm{bath}-N_0^\mathrm{min}}{N_0^\mathrm{max}-N_0^\mathrm{min}} 
\end{align}
\label{ssec:Cs state preparation}

\subsection{Cs temperature}
\begin{figure}
	\includegraphics[width=0.5\textwidth]{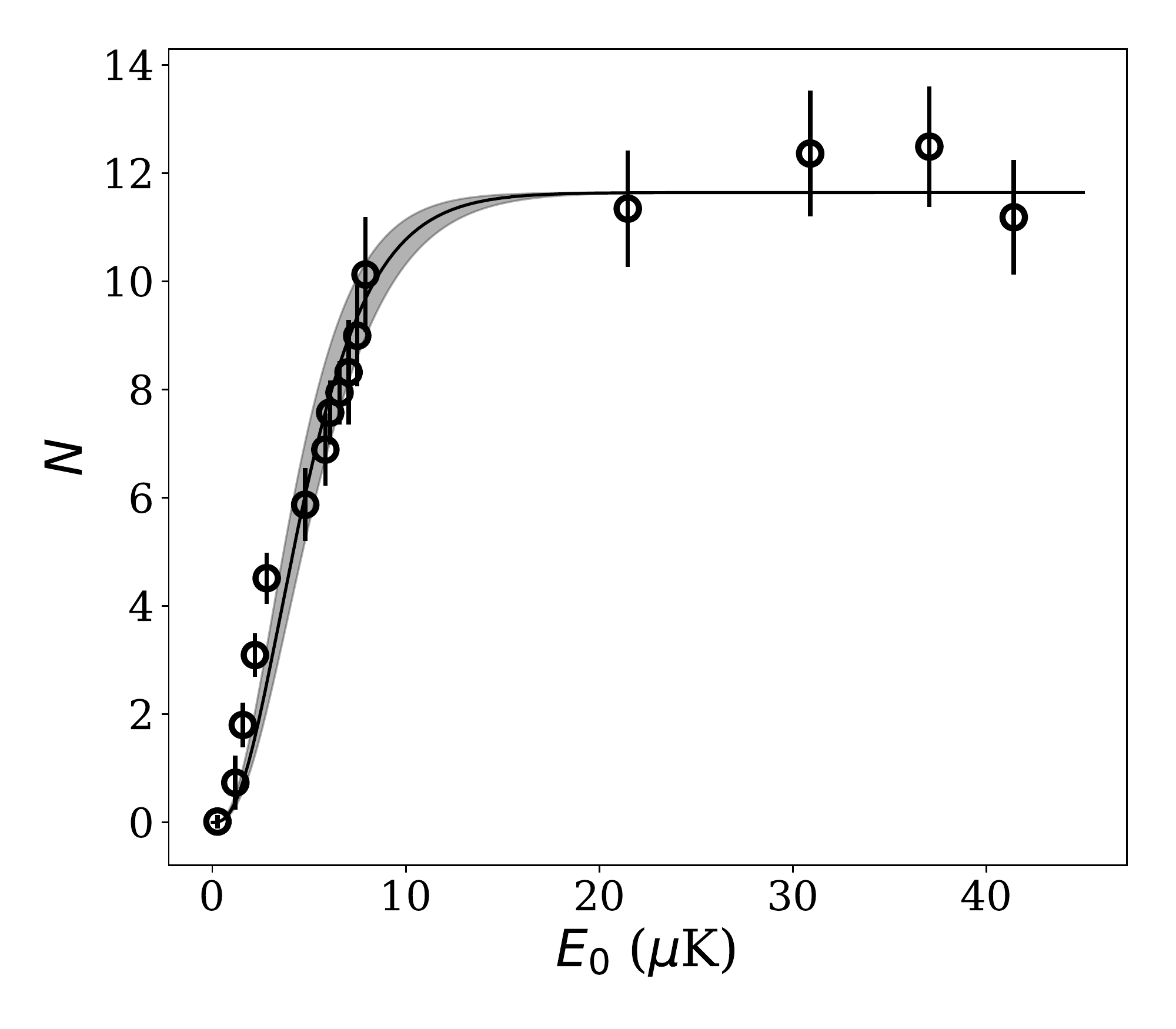}
	\caption{Remaining atoms for the final trap potential $E_0$. The line is the fitted cumulated Maxwell-Boltzmann distribution with $T=1.7\,\mu$K. The shaded area marks the 1$\sigma$ error of the fit.}
	\label{fig:adiabatic lowering}
\end{figure}

We determine the Cs temperature through adiabatically lowering the trap depth and measuring the remaining fraction of atoms \cite{Alt03}. This measurement directly reflects the cumulated energy distribution of atoms in the trap.
Cs is loaded without Rb in the dipole trap potential as described before. Instead of applying the Ramsey sequence,  we adiabatically lower the potential. Atoms with higher kinetic energy leave the trap, while atoms with lower kinetic energy will stay according to the final trap potential. Due to adiabaticity, the final trap depth at which a given class of atoms leave the trap can be related to the original energy of this class of atoms.  Thus by counting the remaining atoms, the initial cumulated in-trap energy distribution can be fitted with a Maxwell-Boltzmann distribution.  For the Cs atoms after Raman cooling,  Fig. ~\ref{fig:adiabatic lowering}  shows the corresponding measurement, where the solid line corresponds to a thermal energy distribution of $T=1.7\,\mu$K.

\subsection{Shift without Rb}
\label{ssec:}
\begin{figure}
	\includegraphics[width=0.45\textwidth]{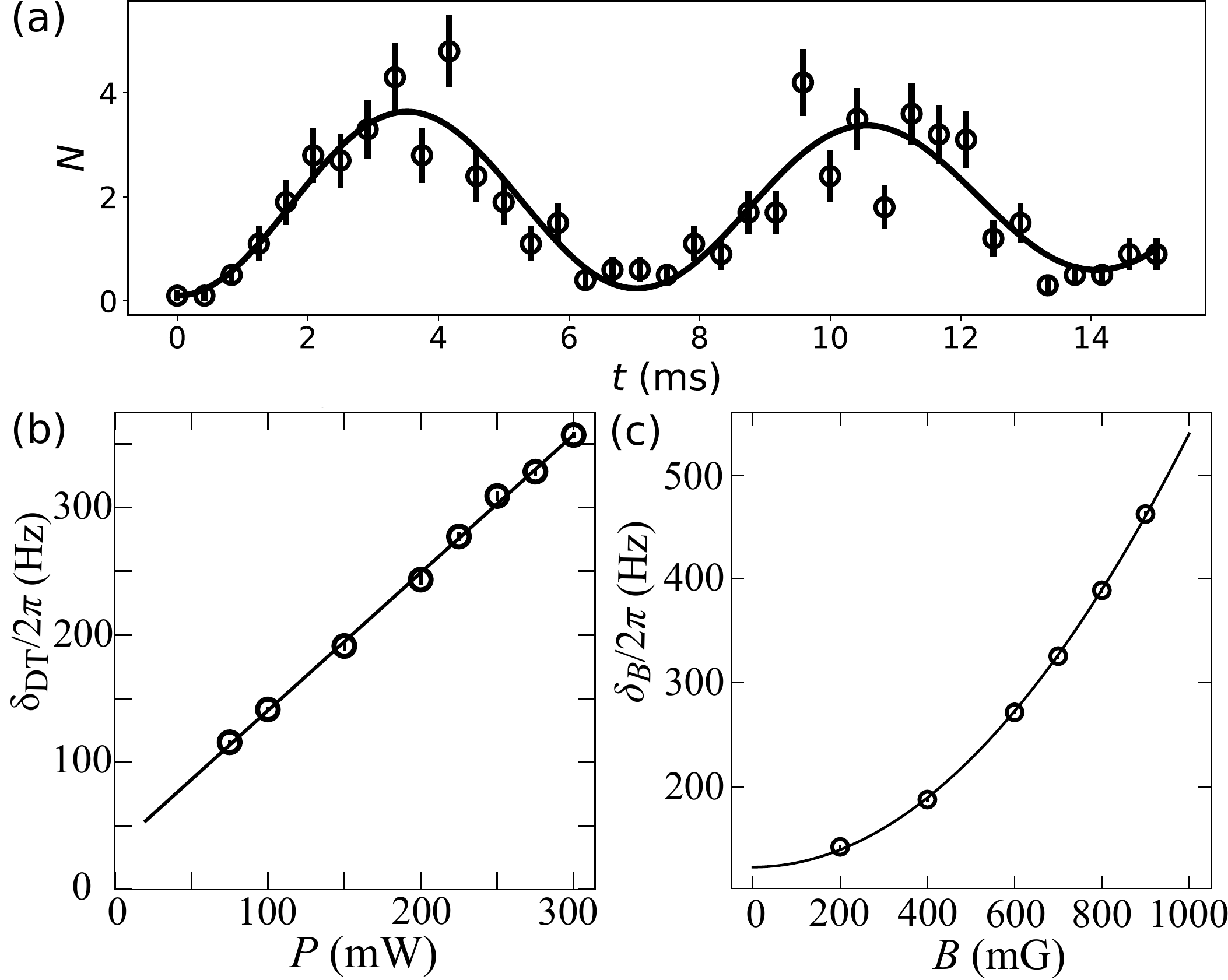}
	\caption{(a) Example of the time evolution of the Ramsey fringe for extracting the detuning.
		(b) Ramsey frequency shift $ \delta $ versus dipole trap power. Data (open dots) are fitted with a linear fit yielding a slope of $\delta_\mathrm{DT}/P = 2\pi \times (1083 \pm 18)$ Hz/W, in good agreement with the theoretical expectation of $2\pi \times 1104$ Hz/W \cite{Kuhr2005}. (c) Ramsey frequency shift $\delta$ versus magnetic field $B$.  The experimental data (open dots) is fitted with a quadratic fit $ \delta = aB^2+c $, where  $a = \delta_B/B^2 = 2\pi \times (417.2 \pm 3.3)$ Hz/G$^2$ in agreement with the theoretical value of $2\pi \times 427.5$ Hz/G$^2$. The offset $c$ is caused by the AC-light shift of the dipoletrap.} 
	\label{fig:shift no Rb}
\end{figure}

The phase shift without Rb comprises the differential AC-Stark shift of the dipole trap potential \cite{Kuhr2005}, and residual Zeeman shifts in the ambient magnetic field. For extraction of the differential light shift, we have measured the time evolution of the $\ket{g}$-state population after the two $\pi/2$-pulses with 0 additional phase $\Phi$ for different trap potentials, see Fig.~\ref{fig:shift no Rb}(a). We fit the time evolution of the detected atom number
with 
\begin{align}
N(t)=0.5 A\left (1-\exp\left[ - \frac{t^2}{T_2^2} \right] \cos(\delta t) \right )+C, 
\end{align}
where $A$ and $C$ are amplitude and offset of the fit, respectively. Since $\delta_\mathrm{DT} \propto U_0$ with $U_0 \propto P$ the dipole trap potential proportional to beam power $P$ we find the expected linear dependence of the shift on power \cite{Kuhr2005} as shown in Fig.~\ref{fig:shift no Rb} (b).

Similar we determine the second-order Zeeman shift with $\delta_B = (g_J-g_I)^2\mu_B^2/(2\hbar\Delta E_\mathrm{hfs})B^2$, with $g_j$ the fine structure Landé $g$-factor and $g_I$ the nuclear $g$-factor, $\mu_B$ the Bohr magneton and $\Delta E_\mathrm{hfs}$ the hyperfine splitting.  Choice of the clock transition suppresses the first-order Zeeman shift. The quadratic behavior of the frequency shift with an applied magnetic field can be seen in Fig.~\ref{fig:shift no Rb} (c), showing good agreement with a theory prediction. \\

\begin{figure}
	\includegraphics[width=0.45\textwidth]{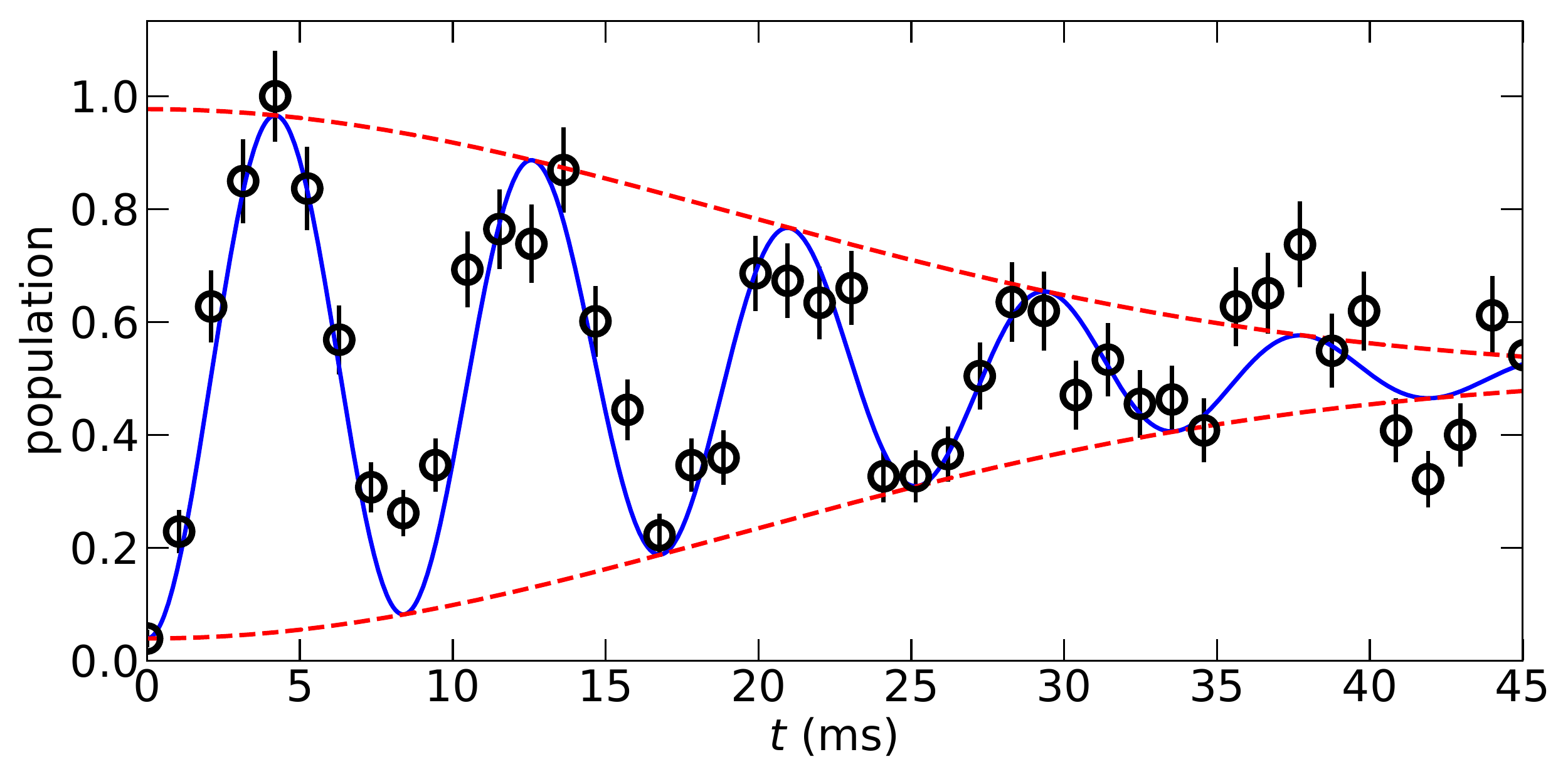}
	\caption{Time evolution of the Ramsey signal for $\varphi=0$\textdegree. The oscillation amplitude decreases with Gaussian envelope, yielding a $T_2$ dephasing time of $T_2 = 27.2\,$ms.}
	\label{fig:T2 no Rb}
\end{figure}

Additionally, we extract the coherence time of Cs atoms without Rb bath from such Ramsey fringe, as shown in Fig.~\ref{fig:T2 no Rb}.  We find a coherence time without Rb of $T_2 = 27.2\,$ms. The decoherence time without Rb can be explained by inhomogeneous differential light shifts originating from the thermal motion of the atoms in the harmonic potential of the dipole trap \cite{Kuhr2005}. However, the dephasing time is roughly one order of magnitude longer than the observed decoherence time in the Rb bath and therefore neglected in the further analysis.

\subsection{B field calibration}
In the regime of the Feshbach resonances, small changes of the magnetic field $B$ result in a large variation of the scattering length $a$. We calibrate the magnetic field via microwave spectroscopy of Rb on the transition $\ket{F_\mathrm{Rb}=1, m_{F,\mathrm{Rb}}=0} \rightarrow \ket{F_\mathrm{Rb}=2, m_{F,\mathrm{Rb}}=1}$. 
The total magnetic field at the position of the atoms $B_\mathrm{exp}=B_\mathrm{coil}+B_\mathrm{bg}$ comprises the field $B_\mathrm{coil}$ generated by coils  as sketched in FIG.~\ref{fig:setup} (a), and the magnetic background field, which causes most unwanted magnetic field changes.
To extract the correct value of $B_\mathrm{exp}$ we apply Microwave radiation  to the atoms, where the frequency $\omega_\mathrm{MW}$ matches the calculated transition frequency of the Rb atoms calculated by the Breit-Rabi formula for the designated magnetic field $B_0 = 198.5$\,mG.
We then vary the magnetic field $B_\mathrm{coil}$,  and hence the transition frequency of the Rb atoms, where we expect the maximum of population transfer for $\Delta B = B_\mathrm{exp}-B_0 = 0\,$mG. 

For a two-level system, the population transferred into the excited state by a near-resonant microwave field is given by
\begin{widetext}
\begin{align}\label{eq:Spectrum}
p(\omega_\mathrm{coil}) = \frac{\Omega_0^2}{\Omega_0^2+(\omega_\mathrm{coil}+\omega_\mathrm{bg}-\omega_\mathrm{MW})^2}\sin^2{\left(0.5\frac{2\pi \sqrt{\Omega_0^2+(\omega_\mathrm{coil}+\omega_\mathrm{bg}-\omega_\mathrm{MW})^2}}{2\Omega_0}\right)}, 
\end{align}
\end{widetext}
with $\Omega_0$ the independently measured bare Rabi frequency for this transition. 
In order to fit the data, we define for the atomic transition frequency $\omega_0 = \omega_\mathrm{coil}+\omega_\mathrm{bg}$  with $\omega_\mathrm{coil}/2\pi = B_\mathrm{coil} \times 0.7 $\,MHz/G and $ \omega_\mathrm{bg}/2\pi = B_\mathrm{bg} \times 0.7 $\,MHz/G assuming linear Zeeman splitting. The linear splitting, rather than the Breit-Rabi formula, is considered for simplicity while not changing the relevant result at $B_\mathrm{exp} = B_0$.
The only free parameter for the fit is $\omega_\mathrm{bg}$. 
For the experiment,  $B_\mathrm{coil}$ is extracted with the condition $\omega_0 = \omega_\mathrm{MW}$ assuring that $B_\mathrm{exp} = B_0$.

\begin{figure}
	\includegraphics[width=0.45\textwidth]{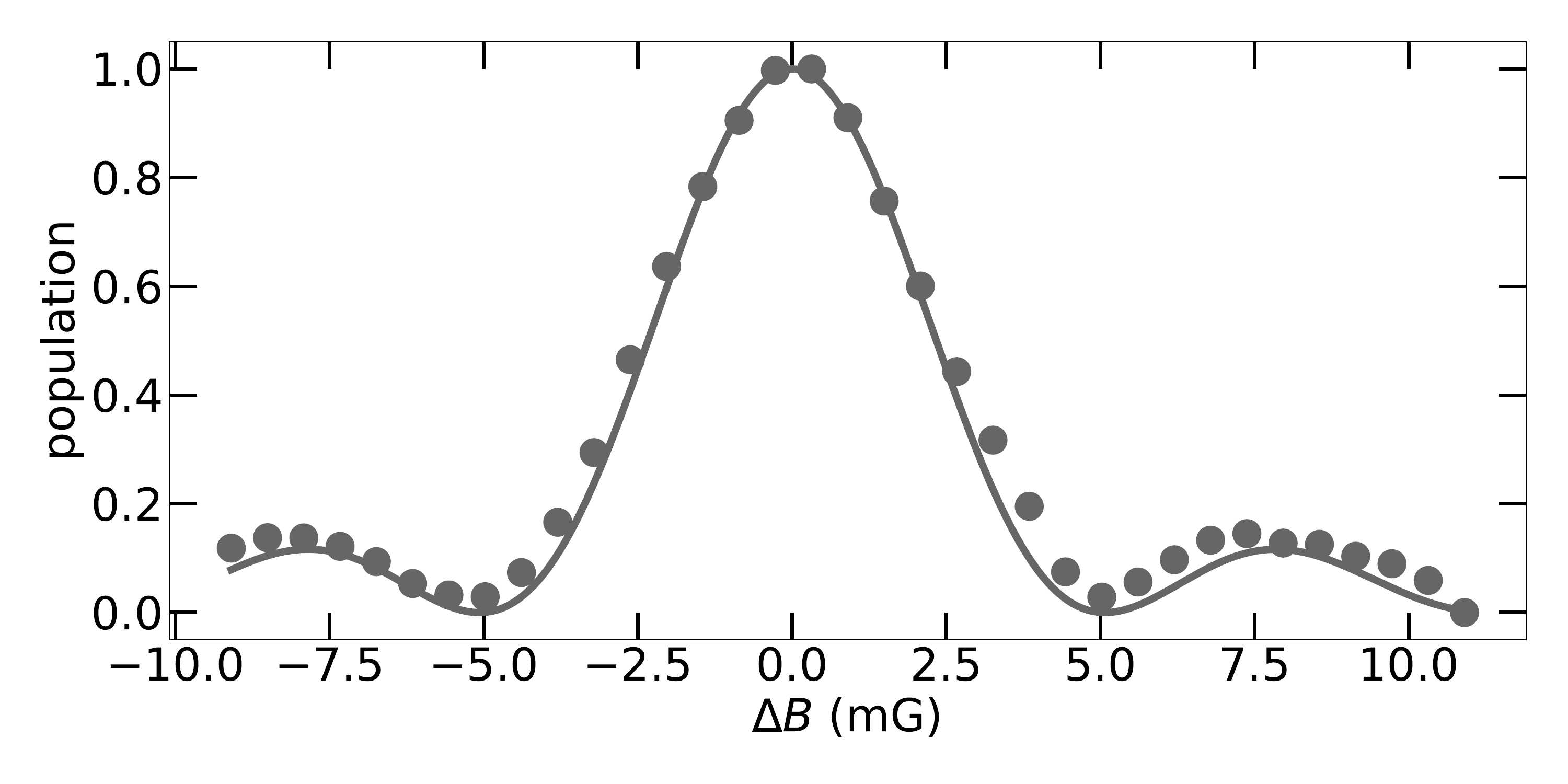}
	\caption{Microwave spectrum of Rb atoms around the magnetic field of $B_0=198.5\,$mG, used throughout this work.}
	\label{fig:B field calibration}
\end{figure}

\subsection{Decoherence}
The decoherence mechanisms are mainly based on the density distribution of the Rb cloud and the thermal average of collision energy $E_\mathrm{coll}$. We numerically integrate the combined average as 
\begin{widetext}
\begin{equation}\label{eq:DephasingModel}
	p(t) = \int_{E_c=0}^{\infty}\left(\int_{V} p_\mathrm{Rb}(\vec{r})\cos^{2} \left(\frac{\delta_\mathrm{Rb}(\vec{r}, B, E_c) t}{2}  + \frac{\varphi}{2} \right) \mathrm{d}V\right)p_\mathrm{MB}(E_c) \mathrm{d}E_c
\end{equation}
\end{widetext}
with 
\begin{equation}
	\delta_\mathrm{Rb}(\vec{r}, B, E_c) = \frac{ 2 \pi \hbar}{\mu} n_\mathrm{Rb} (\vec r) \Delta a (B, E_c),
\end{equation}
where $p_\mathrm{Rb}(\vec{r})$ is the normalized Rb density distribution $n_\mathrm{Rb}(\vec{r})/N$ with $N$ the total Rb atom number, with 
\begin{equation}
	n_\mathrm{Rb}(\vec{r}) = n(0)\exp\left(\frac{m_\mathrm{Rb}\sum_{i}\omega_{i}^2 x_i^2}{2k_B T}\right)
\end{equation}
the Rb density density distribution in a harmonic trap.
$n(0)$ is the peak density, $m_\mathrm{Rb}$ the Rb mass, $\omega_i$ the trap frequency in direction $i={x,y,z}$ and $k_B$ the Boltzmann constant.
\begin{equation}\label{eq:MBD}
p_\mathrm{MB}(E_\mathrm{c},T) = \frac{2\pi}{(\pi k_B T)^{3/2}} \sqrt{E_\mathrm{c}} \exp\left (- \frac{E_\mathrm{c}}{k_B T} \right )
\end{equation}
is a Maxwell-Boltzmann distribution of the collision energies $E_\mathrm{c}$ for temperature $T$. For only changing less than $10$\%, $a_e=539 a_0$, with $a_0$ the Bohr radius, is assumed to be constant for the calculations. 
The Ramsey fringes for different times are fitted with
$ Amp*\cos^2(\frac{\Phi}{2}-\frac{\varphi}{2}) + offset $ to extract the visibility and phase $\Phi$ for each time. The decoherence time $T_2$ is extracted by fitting a Gaussian $\exp(-t^2/T_2^2)$ to the visibility over time $t$. 
To extract the phase evolution we use a linear model for times smaller than $T_2$ time as $\Phi(t)=\delta t + 360 $\textdegree.  

\begin{figure}
	\includegraphics[width=0.45\textwidth]{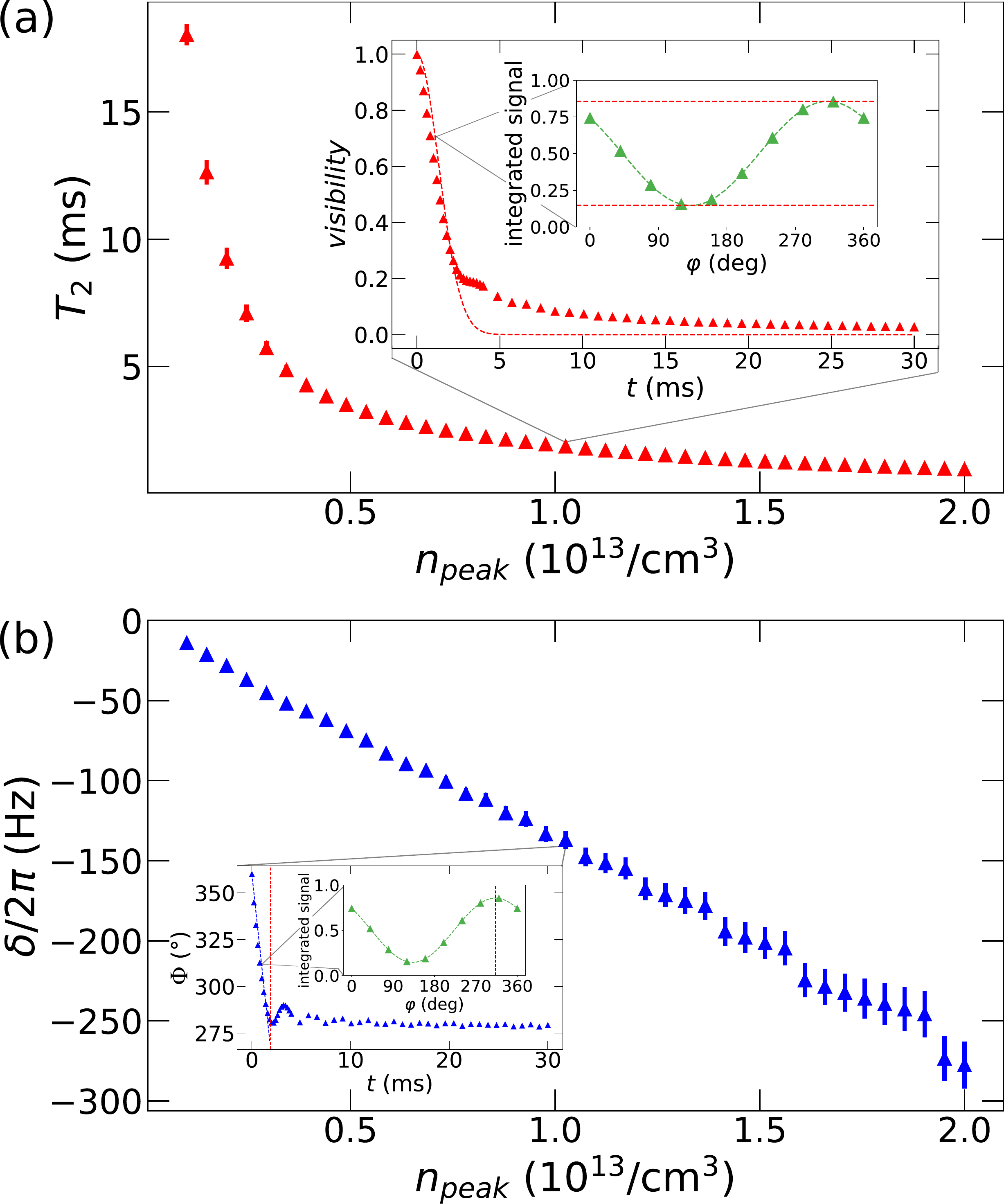}
	\caption{Theoretical extraction of the decoherence time $T_2$ and phase shift $\Phi$. Errorbars (shaded areas in FIG.~\ref{fig:density} and \ref{fig:temperature}) are the standard deviation of the Gaussian and linear fit for extraction of the $T_2$ time and $\delta$, respectively.}
	\label{fig:decoherence theory}
\end{figure}



\end{document}